\pdfoutput=1
\documentclass[10pt]{article}

\usepackage{color}
\usepackage{bbm}
\usepackage{epsfig}
\usepackage{amssymb,amsmath,amscd, mdwmath}
\usepackage{graphicx,cite,url}
\usepackage{algorithmic}
\usepackage{float}
\usepackage{amsfonts}
\usepackage{syntonly}
\usepackage{multirow}
\usepackage{graphicx}
\usepackage{subfig}
\usepackage{tabularx}
\usepackage{setspace}
\usepackage{stfloats}
\usepackage{amsthm}
\usepackage[labelfont=bf,labelsep=period,justification=raggedright]{caption}

\makeatletter
\renewcommand{\@biblabel}[1]{\quad#1.}
\makeatother

\date{}

\begin{document}

\begin{flushleft}
{\Large
\textbf{Modelling the Dropout Patterns of MOOC Learners}
}
\\
Zheng Xie$^{1, \sharp }$
\bf{1}  College of Liberal Arts and Sciences,   National University of Defense Technology, Changsha,   China\\
$^\sharp$ xiezheng81@nudt.edu.cn\\
 \end{flushleft}
\section*{Abstract}
We adopted survival analysis for the viewing durations of massive open online courses. The hazard function of empirical duration data is dominated by a bathtub curve and has the Lindy effect in its tail. To understand the evolutionary mechanisms underlying these features, we categorized learners into two classes due to their different distributions of viewing durations, namely lognormal distribution and power law with exponential cutoff. Two random differential equations are provided to describe the growth patterns of viewing durations for the two classes respectively. The expected duration change rate of the learners featured by lognormal distribution is supposed to be dependent on their past duration, and that of the rest learners is supposed to be inversely proportional to time. Solutions to the equations predict the features of    viewing duration distributions, and those of the hazard function. The equations also reveal the feature of memory and that of memorylessness for the viewing behaviors of the two classes respectively.

\noindent {\bf Keywords:} {Data science applications in education;  Distance education and online learning;   Evaluation methodologies. }

\section*{Introduction}

Massive
   open online courses (MOOCs)
arise from the integration of education and information technologies,   featured by unlimited participation and open access via the Internet\cite{Breslow,ZhuM}. These courses break the spatiotemporal  boundary of   traditional   education,  and contribute to balancing  the resources of education\cite{Emanuel,Reich}, and so on. Differences between these courses and traditional courses lie in several dimensions, involving admission condition, learning motivation, teaching methods, the management of learners, the interactions between teachers and learners\cite{Anderson,Jona2014}. Analyzing learning behaviors has become a hot topic in the MOOC community, which includes the  achievements of learners\cite{DeBoer,Meyer}, the interactions among learners\cite{Sunar}, the visual analysis of course data\cite{Emmons}, the assessment of courses\cite{Sandeen}, and so on.

High dropout rate is a feature of learning MOOCs, which has been regarded as a result of the diversified expectations and motivations to learn these courses\cite{Freitas,Greene,Hone,Littlejohn}.  The learners of MOOCs are motivated not just to pass exams or to obtain certificates. They could be only interested in understanding particular concepts or some   contents\cite{Barak,Barba,Watted,Zheng}, and then are likely to drop after obtaining what they want. Analyzing the dropouts of MOOCs contributes to quantifying the completion and continuance of learning\cite{Alraimi,Jordan}. For example, how many learners who will continue to learn when they have learned for a certain time? At what rate they will drop in the future? Moreover, understanding the evolutionary mechanisms underlying the dropout behavior and modelling them mathematically contribute  to profiling learners' type\cite{Kizilcec1},  and to quantifying the effects of teaching methods or other explanatory variables on learning behaviors\cite{Hew2016,LiLY,Cheng}.


The log data of learning behaviors collected by the platforms of MOOCs  can be used to analyze the dropout rate, where viewing is a prominent learning behavior\cite{Henrie}. We adopted survival analysis for learners' viewing duration  on  a course, where the empirical data are provided by the platform iCourse (http://www.icourse163.org). The  survival function of the durations describes the fraction of the learners   viewing a course past a given time, and the corresponding hazard function describes the rate of these learners   dropping viewing at the given time. The hazard function of empirical data is dominated by a concave function, but decreases in the tail. The shape of the concave part  is known as bathtub curve\cite{Klutke}, and the decreasing tail part is known as the Lindy effect\cite{Holman}, namely the more you have learned, the more you want to learn. The features of the hazard function  have the potential to be utilized  to assess courses' attraction.

We analyzed the evolutionary mechanisms underlying these features. We categorized learners as lognormal-learners and segment-learners  due to the different features of their viewing duration distribution, namely lognormal distribution and power law with exponential cutoff. For each category, we provided a random differential equation to describe its growth mechanism of viewing durations. For lognormal-learners, we assumed that their duration change rate correlates to their past duration and to a random disturbance. For segment-learners, we assumed  that their duration change rate is inversely proportional to time, and that their start time of viewing a course is a random variable. The equations express the factor of memory in viewing behavior for lognormal-learners, and that of memorylessness for segment-learners. The solutions to the equations reproduce the features of  the density function and those of the hazard function.

This paper is organized as follows. Empirical  data are described in Section  2. The    hazard function of viewing durations  is described in Section 3. The evolutionary mechanisms  of the  hazard function  are analyzed  in Section 4-6. Discussion and conclusions are drawn in Section 7.

\section*{The data }
\label{s:Experimental}
\noindent

We analyzed the log data of viewing eight courses from 01/01/2017 to 10/11/2017, which are provided by the platform iCourse. The courses are selected from natural sciences, social sciences, humanities and engineering technologies respectively. Specific statistical indexes of these courses are listed in Table~\ref{tab1}, which have been used to analyze course attraction in our previous work\cite{Xie7}. The data include the time length of each video. For each learner, the data include the start time of viewing  each video he or she opened, and
  the corresponding viewing durations.

Videos can be downloaded by iCourse app. The log data of viewing the downloaded videos are also collected, unless the app disconnects to the Internet. Accordingly, our study only involves the online viewing behavior, recorded as log data. However, learners may be off-task during video playing, which cannot be measured through log data. This is a limitation of our study. In addition, some typical operations on videos are   not analyzed here, such as pausing, skipping, backward, forward, speed changing, and so on.

 We concentrated on learners' viewing duration, which   is defined to be
  the   duration of video playing, where   the duration   of pause is not   counted. We introduced  the following symbols  to express the duration.
    Suppose that  learners $\{L_1,...,L_m\}$ have viewed    a course  with $n$ videos $\{V_1,...,V_n\}$.
 Denote the time length of video $V_i$   as $l_i$,
  the duration of learner $L_s$ viewing  $V_i$ as $t^s_{i}$. Then   the viewing duration of learner $L_s$ on the course is
  $\sum^n_{i=1}t^s_{i}$. Hereafter, the viewing duration on a course  is called   duration for short.

\begin{table*}[!ht] \centering \caption{{\bf Specific statistical indexes of   the  data provided by{ iCourse}. } }
\vskip 3mm
\footnotesize\begin{tabular}{l cccccccc } \hline
 Course  & Course Id& $m$ &	$n$&  	$a$ &$ b$ & $c$ & $d$ \\ \hline
{\it Calculus}&1002301004&	2,955&	     129&	2	&8.081&	0.998&	0.189\\
{\it Game theory}&	1002223009&4,764&	38	&66&7.141	&2.238&	0.427\\
{\it Finance	}& 1002301014  &6,380	&63	&2&5.368&	1.310&	0.330\\
{\it Psychology}&	1002301008&3,827	&26	&59&5.008	&0.913&	0.204\\
{\it Spoken English}& 1002299019	&11,719&	46	&	7&3.032&	0.321&	0.106\\
{\it Etiquette}&1002242007	&3,846&	41	&	22&7.787	&1.271&	0.205\\
{\it C Language}&1002303013 &	17,541&	81	&	39&12.47&	1.541&	0.142\\
{\it Python}&	1002235009&13,417&	53	&	28&10.32&	0.896&	0.087\\
\hline
 \end{tabular}
  \begin{flushleft}
    Index  $m$: the number of learners, $n$: the number of videos, $a$: the number of all-rounders
  who viewed all of the videos,
 $b$: the average number   of    viewed videos per  learner,
  $c$:   the average viewing duration  of learners  (unit: hour),  and $d$: the average time length	of videos (unit: hour).
\end{flushleft}
\label{tab1}
\end{table*}

\section*{The survival analysis of   viewing durations}
\noindent

A learner's viewing duration    on a course
can be regarded as the  ``lifetime" of  his or her  viewing behavior.
The number of learners with duration $t$  expresses   the number of
dropouts at ``age" $t$, where $t\in [0,T]$, and $T$ is the maximum viewing duration.
Therefore, the   density function of viewing durations, denoted as $f(t)$,   expresses the rate of dropouts at any possible  $t$.
 The rate of dropouts at a given  $t$
  for the learners with viewing duration no less than $t$
   is calculated as
   $h(t)=f(t)/S(t)$, where   $S(t)= \int^T_tf(\tau)d\tau$
is the probability of a learner's duration no less  than   $t$.
 In survival analysis\cite{Kleinbaum},   $S(t)$ and   $h(t)$
are called survival function and  hazard function respectively.
The function $h(t)$    is
the derivative of $\log S(t)$,   and then is   more informative about dropouts.

\begin{figure*}\centering
\includegraphics[height=2.8   in,width=6   in,angle=0]{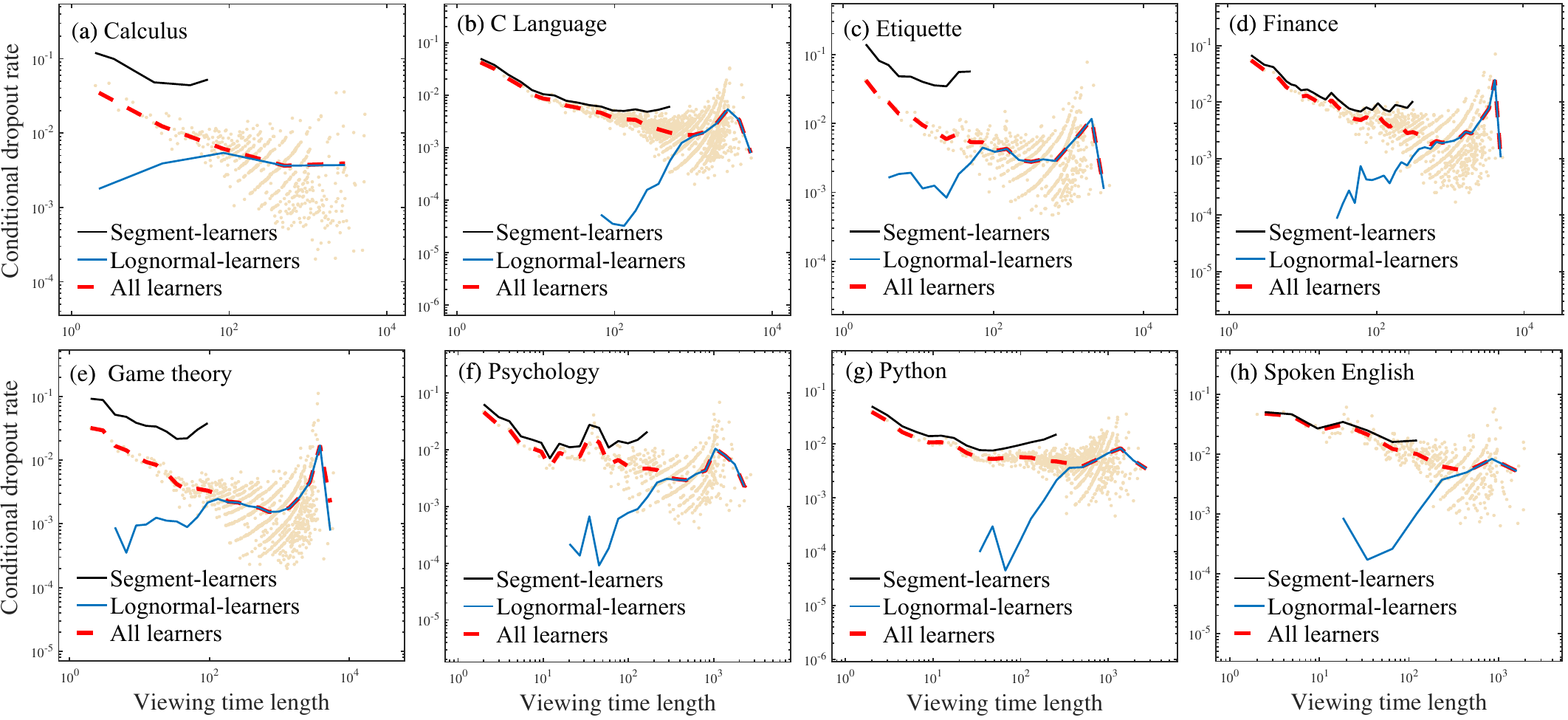}
\caption{  { \bf  The  hazard
functions of empirical viewing durations. }Panels show  the hazard functions  (orange circles) and their trend line (red dotted lines) of viewing durations.  Panels also show the   trend lines of the hazard functions for lognormal-learners (blue lines) and  segment-learners (black lines) respectively.
 } \label{fig1}      
\end{figure*}

The hazard function of empirical data is dominated by a concave function and has a decreasing tail~(Fig.~\ref{fig1}). The shape of the concave part  is known as   bathtub curve, the concept of which comes from product quality assessment. The curve is often used to describe the failures of products over time, which contain the  decreasing rate of early failures as defective products are discarded, the random failures with a constant rate during the useful life of products, and the increasing rate of wear-out failures as the products exceed their designed lifetime. In this study, we called the dropouts of the increasing part    wear-out dropouts.
The   decreasing tail is known as the Lindy effect, namely the future life expectancy is proportional to their current age. It means that every additional period of duration implies a longer remaining duration expectancy.

\begin{figure*}\centering
\includegraphics[height=3.   in,width=6   in,angle=0]{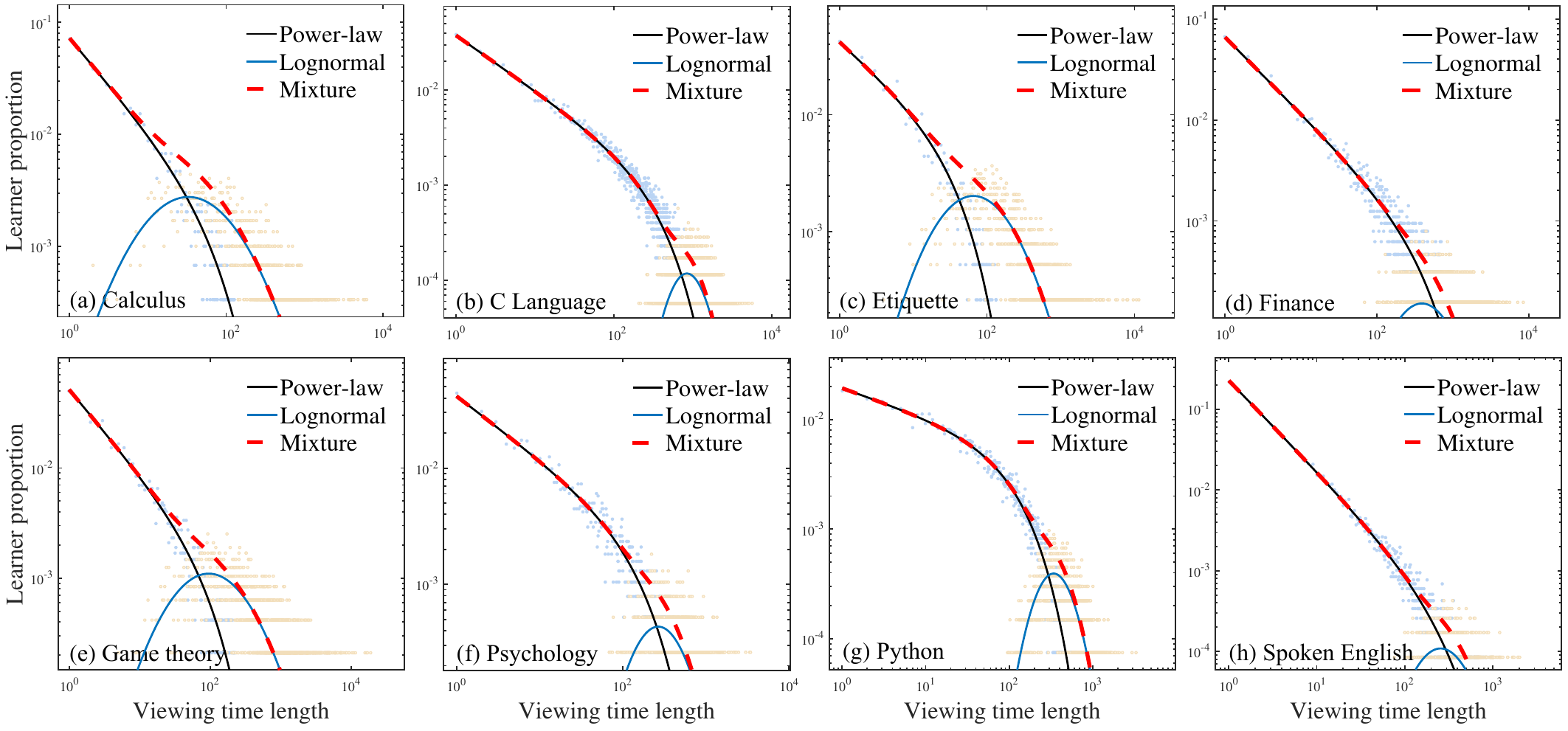}
\caption{  { \bf   Synthetic viewing duration distributions. } Panels show  the
density functions   of lognormal distributions (blue lines,
orange circles)  and those of  the  power-law distributions that have  an exponential cutoff  (black lines, blue circles) respectively, as well as the mixture density functions of them (red dotted lines). The parameters of these distributions are listed in Table~\ref{tab2}. } \label{fig2}      

\end{figure*}
\begin{figure*}\centering
\includegraphics[height=3.   in,width=6   in,angle=0]{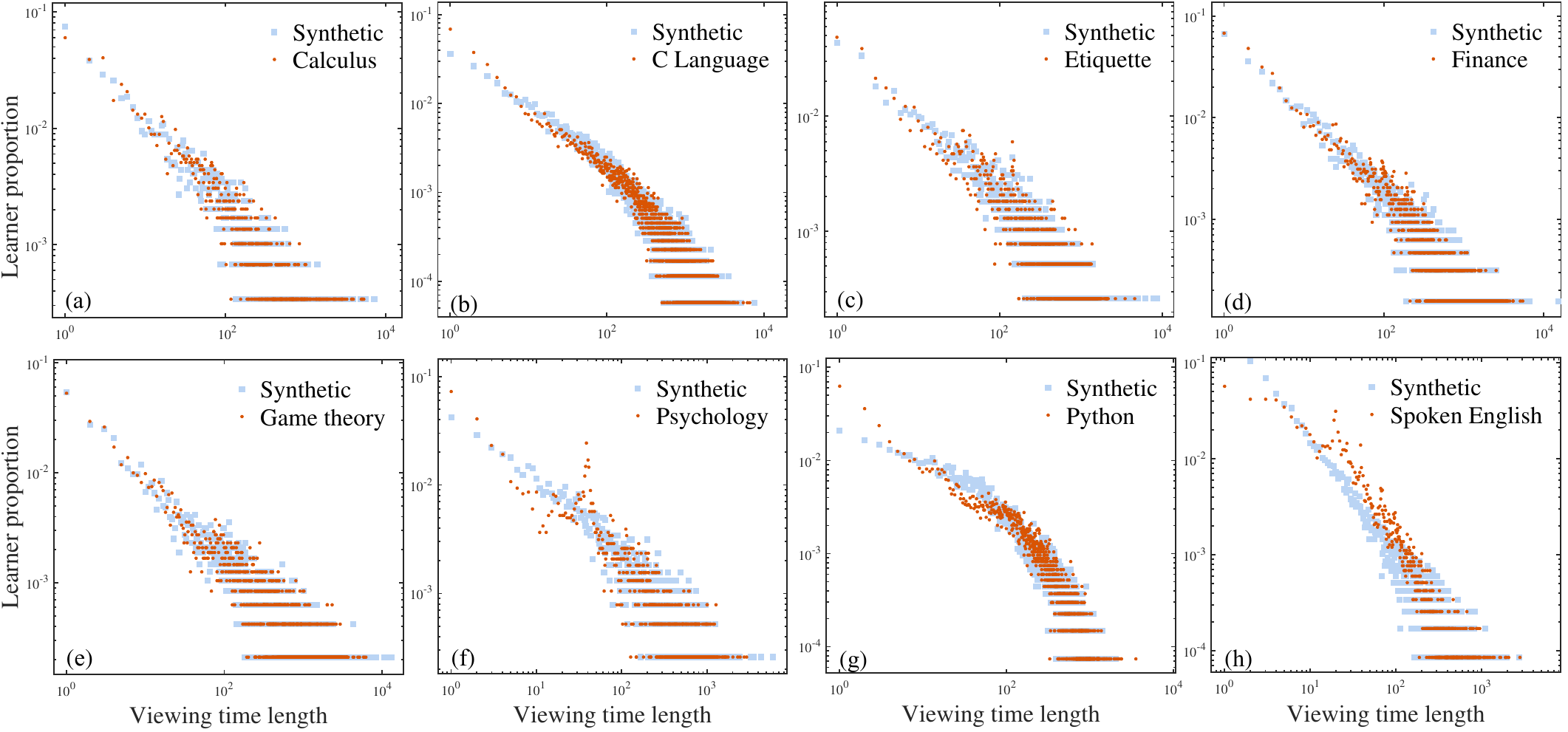}
\caption{  { \bf   Comparisons  between  the empirical distributions of viewing durations and   synthetic ones. } Panels show
that each  empirical  distribution can be approximated by a mixture of
a lognormal distribution and a power law with an exponential cutoff.
} \label{fig3}      
\end{figure*}

To understand why the Lindy effect and bathtub curve  emerge simultaneously,
we should analyze the features of  the density   function $f(t)$, and mine the dynamic mechanisms under those features, because
  the
hazard function $h(t)$ is       based on  $f(t)$.
When  a learner has  viewed  a certain number of videos, his or her viewing duration follows a lognormal
distribution (the results of the KS test are shown  in Table~\ref{tab2}).
For the rest learners, most of their duration follow
a power law   with an exponential cutoff (the results of   good-of-fit   are listed in Table~\ref{tab2}).
To illustrate these features,
we fitted the parameters of these distributions for each course,
 and generated
 synthetic durations following each distribution~(Fig.~\ref{fig2}), the number of which is the same as that of the corresponding
 empirical durations. The comparison  between    empirical duration distributions and    synthetic ones  are shown in Fig.~\ref{fig3}.
What is the relationship between these features of the density function and  the shape of  the hazard function?
To  answer this question,  we explored
the evolutionary mechanisms underlying these features   in the following sections.

\begin{table*}[!ht] \centering \caption{{\bf Fitting  parameters and    goodness-of-fit. } }
\vskip 3mm
\footnotesize\begin{tabular}{l cccccccccccc } \hline
 Course  &   $\alpha$ &	$\beta$&	   $\mu$ & $\sigma$& $\tau$ &p-value & $\psi$\\ \hline
{\it Calculus}&  0.7284  &  0.0035 &  6.1694  &0.8155  & 11 &0.059   & 15.40\%\\
{\it C Language}&0.5803  &  0.0028  &  6.9664   & 0.4865   &29&0.067 &12.48\%\\
{\it Etiquette}& 0.5288 &  0.0151 &  5.5873  &  0.9395    &4& 0.069 & 16.09\%\\
{\it Finance	}& 0.7557  &  0.0024&    6.6165 &   0.8052  &8& 0.092 & 14.53\%\\
{\it Game theory}&	0.6556 & 0.0023   &6.7109    & 0.7546 &7&0.105 &  16.32\%\\
{\it Psychology}&	0.5391 &   0.0062 &   6.0323  &  0.6659  &7& 0.355 & 23.05\%\\
{\it Python}&	 0.2622  &  0.0083  &  6.0817  &  0.4989  &19& 0.077 &16.08\%\\
{\it Spoken English}&0.9561  &  0.0044  &  5.9030 &   0.5965 &10&  0.123  & 30.36\%\\
\hline
 \end{tabular}
  \begin{flushleft} The parameters of  $x^{\alpha}\mathrm{e}^{\beta x}$ are fitted through multiple linear regression.
  The parameters of Lognormal$(\mu, \sigma)$ are   calculated based on   empirical data.  At significance level 5\%, the  KS test cannot reject  that the viewing durations of  the learners, who  have viewed
  no less  than   $\tau$
  videos, follow a lognormal distribution ($p$-values$>0.05$). The good-of-fit index $\psi$ is the half of the cumulative difference  between the duration distribution of  segment-learners
  and the corresponding synthetic one.
 \end{flushleft}
\label{tab2}
\end{table*}

\subsection*{The Lindy effect, wear-out
dropouts  and lognormal distribution}
\noindent
 The empirical data show that the  viewing durations of the learners,  who have viewed no less than $\tau$ videos~(Table~\ref{tab2}), follow a lognormal distribution.  We  called them    lognormal-learners,  and used an algorithm in Reference \cite{Xie7} to find   them (Table~\ref{tab4} in the appendix).
Following
 an identical lognormal distribution
  means those learners are the samples   drawn from the same population in the sense
  of viewing duration; thus   can be categorized as one class.

\begin{figure*}\centering
\includegraphics[height=3.   in,width=6   in,angle=0]{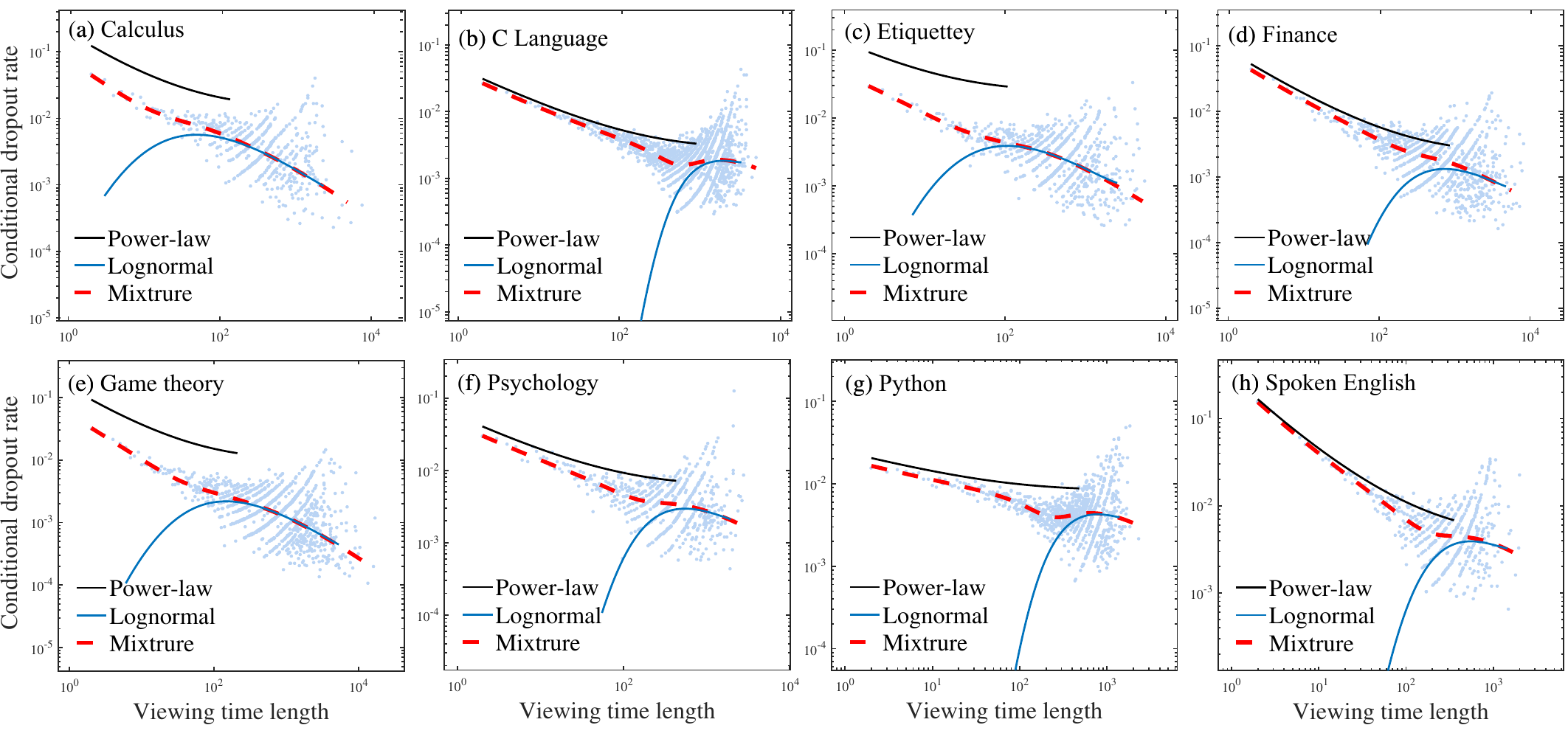}
\caption{  { \bf The  hazard
functions of synthetic viewing   durations.  }Panels show  the
hazard functions   (blue squares)    and their trend line  (red dotted lines) of synthetic durations. Panels also show
the hazard functions of  lognormal  distributions (blue lines) and those of the power-law distributions that have  an exponential cutoff (black lines).
 } \label{fig4}      
\end{figure*}

The density function of lognormal distribution  is $f(x)=\mathrm{e}^{\frac{1}{2}\left(  \frac{\log x-\mu}{\sigma}  \right)^2}/ { \sigma \sqrt{2\pi}x}$, where $x\in [1,  \infty)$, $\sigma>0$, and
$\mu\in \mathbb{R}$.
The corresponding hazard function  is
\begin{align}\label{eq1}h(x)= \frac{1}{x\sigma  }\sqrt{\frac{2}{ \pi}} \mathrm{e}^{-\frac{ (\log x-\mu)^2 }{2\sigma^2}}\left(1-\mathrm{erf}\left(\frac{\log x-\mu}{\sqrt{2}\sigma}\right)\right)^{-1}.\end{align}
The hazard function (\ref{eq1}) is convex, when its corresponding density function  is convex\cite{Kiefer}.  Fig.~\ref{fig1}, \ref{fig2} show      that the density function of   lognormal-learners and  the corresponding hazard function   are convex   for each empirical course,  namely
     appears
the wear-out
dropouts and the Lindy effect.  Fig.~\ref{fig4} shows that the hazard functions of   synthetic durations  give
  reasonable fits to those of empirical data, which  verify above  arguments.

To find the evolutionary mechanisms underlying the wear-out
dropouts and the Lindy effect, we should go back to where   lognormal distributions come from.
 They often emerge   in the  lifetime distributions of mechanism units\cite{Gaddum}, where the   lifetime of a unit is affected   by
 the multiplication of   many small factors.
 Approximate  these factors as  a range of   independent and identically distributed  random variables.
  The central limit theorem says that  the summation  of these variables in log scale follows a normal distribution.
Transforming   to the original scale gives rise to  that the
multiplication of these factors follows a lognormal distribution.
This is   known as  the multiplicative version of the central limit theorem, called the Gibrat's law\cite{Sutton}.

Lognormal-learners  contain the all-rounders  who viewed all of the   videos.
It means that   the endurances of the rest lognormal-learners    and those of
   all-rounders are homogenous; thus
     could be regarded as potential all-rounders.
For a potential all-rounder who has the willing  to complete a course, his endurance   of viewing   videos (measured by his viewing duration)   could be analogized  to a mechanical unit   whose failure mode is of a fatigue-stress nature. The life of such a unit follows a lognormal distribution.
 The analogy enlightens us to provide
\begin{align}\label{eq2}  {d}  k_i(t)= \mu k_i(t) dt+\sigma k_i(t) dw,\end{align}
where $k_i(t)$ is the duration at time $t$ of  learner $i$,
$w$ is     the Wiener process,  $\mu$ and $\sigma>0$.  Supposing  $k_i(t_0)=1$ gives rise to
the solution $k_i(t)=\mathrm{e}^{(\mu-\sigma^2/{2}) (t-t_0)+ \sigma \int^t_{t_0}dw},$ which
is the random variable of a lognormal distribution.

Eq. (\ref{eq2})   means
the viewing behavior of a lognormal-learner  has memory, because the  change rate of duration       correlates to his or her past duration.
Moreover,
the expected   change rate  correlates to the past duration positively. It means when $t$ is large enough, the duration  increases  exponentially, namely
the more you learn, the more you want to learn. This is the status of the learners who are deeply impressed by a course.

\section*{Dropouts with a constant rate  and exponential cutoff}
\begin{figure*}\centering
\includegraphics[height=3.   in,width=6   in,angle=0]{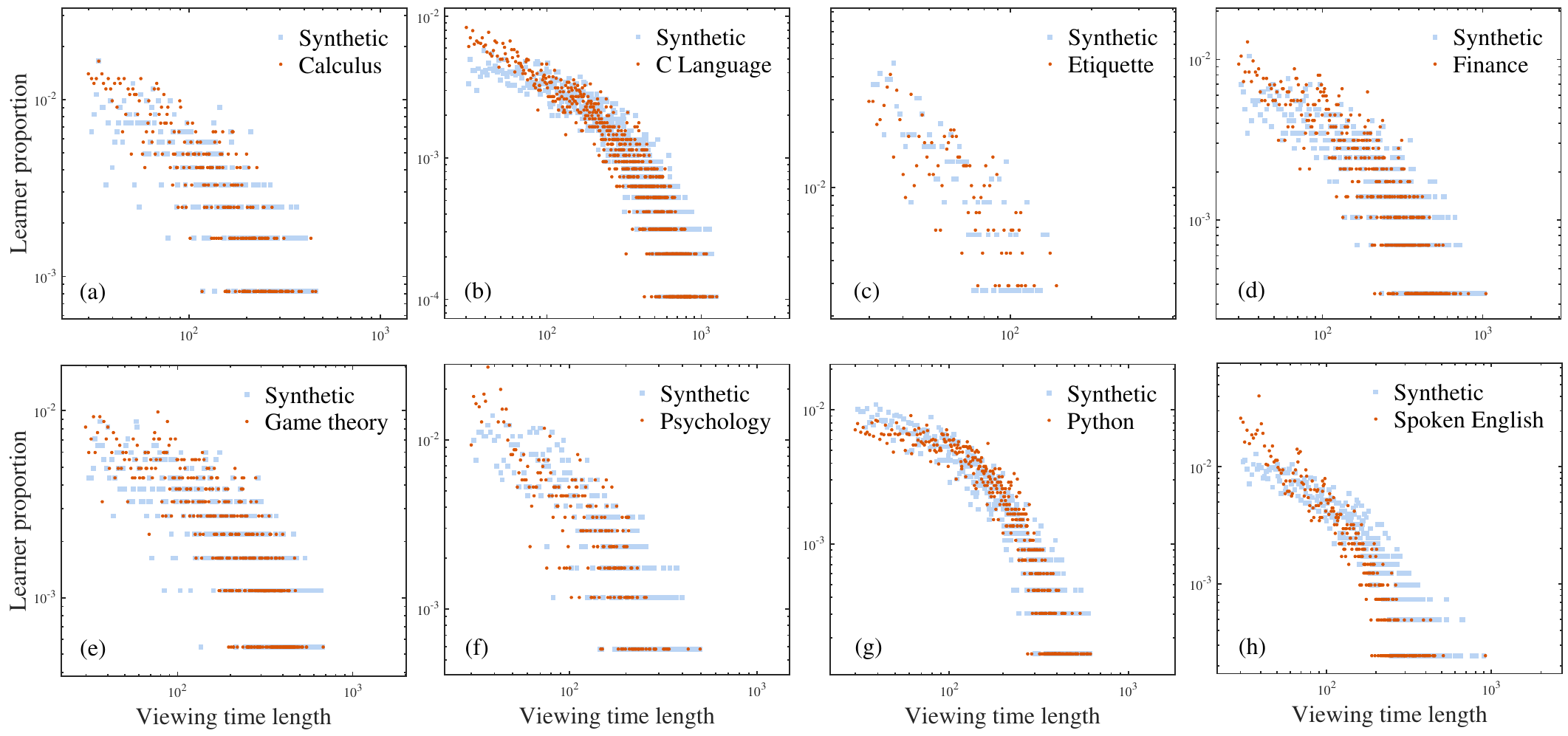}
\caption{  { \bf The exponential cutoffs of viewing duration distributions.}  Panels show the duration  distributions
   of the segment-learners that viewed   no less than one minute (red circles),   compared with  the predictions of Eq.~(\ref{eq3}) (blue squares).  }  \label{fig5}      
\end{figure*}

\noindent
Fig.~\ref{fig3} shows the viewing durations of the learners
viewing less than $\tau$ videos (Table~\ref{tab2})
  approximately  follow  a power law   with an exponential cutoff.
  The emergence of   the cutoff   is mainly due to that the durations  of segment-learners,  which are no less than one minute,   approximately follow an exponential distribution~(Fig.~\ref{fig5}).
The density function of exponential distribution is  $f(x)=\mathrm{e}^{- {x}/{\lambda}}/{\lambda},$ where $x\in[1,\infty)$  and $\lambda>0$.
The corresponding survivor function is
 $S(x)= \mathrm{e}^{- {x}/{\lambda}}$, and then the   hazard function is a constant, namely  $h(x)={f(x)}/{S(x)}={1}/{\lambda}$.

To find the evolutionary mechanism  underlying the dropouts with a constant rate, we  came back to the mechanism underlying  exponential distribution. The distribution is featured by memorylessness, because it satisfies
$p(T>s+t|T>s)={\mathrm{e}^{-(s+t)/\lambda}}/{\mathrm{e}^{-s/\lambda }}=\mathrm{e}^{- t/\lambda}=p(T>t)$ for any possible $T$, $s$,  and $t$.
In our study, the memorylessness   means the future   viewing duration is free of  the past duration.
For example, the  probability that
a learner, who has viewed ten minutes, will view one  minute is equal to the  probability that a learner, who dose not view videos, will view one minute.


Due to the memorylessness and the fatigue of learning,
it is reasonable to suppose  the change rate of viewing duration  decreases with  time. For simplicity, we supposed   the change rate of   learner $i$'s   viewing duration $k_i(t)$ to be
\begin{align}\label{eq3} \frac{d }{d t}k_i(t)= \frac{\lambda}{t} .\end{align}
Solving the equation in the time interval $[y,T]$  gives rise to
\begin{align}\label{eq4}  k_i(T)  =  {\lambda}\log{\frac{T} {y} },  \end{align}
 where  $y$ is  the start time of viewing.
 Supposing
$y$ is a random variable of the uniform distribution over $[T_0, T]$ gives rise to
\begin{align}\label{eq5}p(k_i\leq x)= &p({\lambda}\log{\frac{T}{y}} \leq x) = p({T}{ \mathrm{e}^{-\frac{x}{\lambda}} }<  {y} )
\notag\\=&1-   \frac{T}{T-T_0}  { \mathrm{e}^{-\frac{x}{\lambda}} }.\end{align}
It leads to an exponential distribution
\begin{align}\label{eq6}p(k_i=x)=\frac{d}{dx} p(k_i\leq x) =  \frac{T}{\lambda(T-T_0)}  \mathrm{e}^{-\frac{x}{\lambda}} . \end{align}
When $T_0=0$, Eq.~(\ref{eq6}) is the standard exponential distribution.

To make synthetic duration distributions fit the empirical ones, we valued the parameters of the solution~(\ref{eq4}) based on the
information of empirical data.
Let the domain  of
 the durations of segment-learners (which  are no less than one minute)
 be $[T_1,  T_2]$, and  calculate the exponent $-1/\lambda$  of the formula (\ref{eq6}) by fitting  empirical data (Table~\ref{tab3}).
Letting the simulated duration ${\lambda}\log{ ({T}/ {y} )}$ belong  to $[T_1, T_2]$  gives the sampling interval $[\mathrm{e}^{-T_2/\lambda}, \mathrm{e}^{-T_1/\lambda} ]$ for   $y/T$.
Table~\ref{tab6} in the appendix shows the detail of this simulation process.

 Analytical arguments allow for the prediction
of the exponential cutoff. The simulations based on the solution (\ref{eq4}) also provide a reasonable fit to those of
empirical data~(Fig.~\ref{fig5}, the index     $\psi_1$ in Table~\ref{tab4}).
 Therefore, Eq.~(\ref{eq3}) can be regarded as an  expression of the evolutionary mechanism for  the exponential cutoff and for the random dropouts with a constant rate.




\begin{table*}[!ht] \centering \caption{{\bf  Parameters of   synthetic   power law and    exponential cutoff. } }
\vskip 3mm
\footnotesize\begin{tabular}{l rrrr|rrrrrrrrrr } \hline
 Course  &  $T$ &   $\lambda$ & $N_1$  &$\psi_1$&	$a$ &  $b$ &$c$&$N_2$& $\psi_2$  \\ \hline
{\it Calculus}&460&   116.9& 1,217  &26.38\% &1.30e03 &3.682 &7.25e-2&1,169 &   11.89\%\\
{\it C Language}& 1,253 &224.5 & 9,587 &16.27\% &8.30e03 &2.383&    7.21e-2 &5,587 &  8.5\%\\
{\it Etiquette}&206& 42.62 & 683 &22.36\% &1.89e01&2.122 &7.71e-2&  1,079& 10.29\%\\
{\it Finance	}&1,042  &178.1&2,875 & 24.59\%&3.98e03 & 4.093 &6.98e-2& 2,448 & 7.52\%\\
{\it Game theory}&678& 181.4&1,834&25.08\%  &1.47e02 &2.904&7.51e-2&1,408& 7.74\%\\
{\it Psychology}&495&  93.70 &1,718 &31.49\%  &1.04e01& 2.170 &5.54e-2& 1,188&17.51\%  \\
{\it Python}& 616& 104.2&  6,607&14.59\%&2.70e00 &1.351&6.94e-2 &4,261 &13.49\%  \\
{\it Spoken English}&  916&93.70&4,062&25.58\%& 1.29e36&22.78 & 5.92e-2&7,074& 18.52\%\\
\hline
 \end{tabular}
  \begin{flushleft}
  Index $T$: the maximum duration of segment-learners, $\lambda$: the parameter of Eq.~(\ref{eq3}),
   $N_1$ and  $\psi_1$ ($N_2$ and  $\psi_2$): the number  of the segment-learners with duration   no less than (less than) one minute and the  half of the cumulative
 difference between the duration distribution   of those learners and the corresponding  synthetic  distribution,
 $a$ and $b$: the parameters of Eq.~(\ref{eq8}), $c$: the normalization coefficient of Formula~(\ref{eq7}).
 \end{flushleft}
\label{tab3}
\end{table*}

\subsection*{Early  decreasing  dropouts and  power law }
\begin{figure*}\centering
\includegraphics[height=3.   in,width=6   in,angle=0]{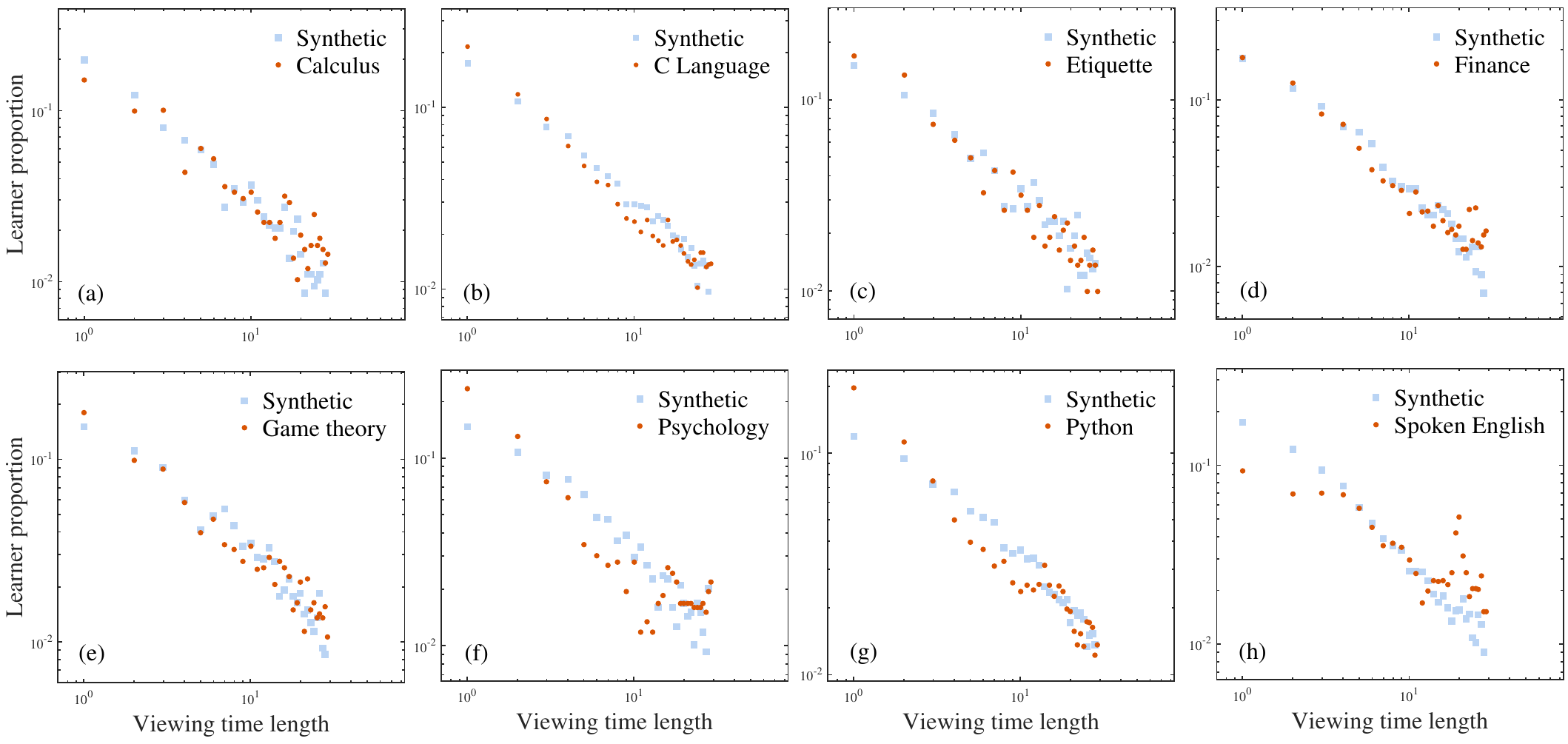}
\caption{   { \bf The   power-law parts   of   viewing duration distributions.}  Panels show   the
 duration distributions of the  segment-learners who viewed      less than one minute (red circles),   compared with the  predictions of Eq.~(\ref{eq9}) (blue squares).  }
 \label{fig6}      
\end{figure*}

\noindent
The early decreasing trend  of dropout rates appears in the hazard functions of empirical data, which  describes a phenomenon that the dropout rate of viewing course
decreases   within the first minute.
 It describes the  period of ``infant mortality" where
 the learners, who only tour a course,   drop viewing.
 Meanwhile, the density functions of   empirical data show the viewing durations of segment-learners, who viewed     less than one minute, can be fitted by  a power-law    function approximately~(Fig.~\ref{fig6}).
Denote the density function of a power-law distribution  by  $f(x)=cx^{-\alpha}$, where $c$ is the normalization coefficient,  and $\alpha\in(0,1)$. Hence the random variable $x$ is valued in a finite interval, denoted as $[R_1, R_2]$.
 Note that the value of the exponent $\alpha$ is different from that of degree distribution in network sciences, which is larger than one.
The corresponding  survivor function   is
 $S(x)= 1-   {c}(x^{1-\alpha}-1)/({1-\alpha})$, and  the   hazard function is
 \begin{align}\label{eq7}h(x)= \frac{cx^{-\alpha}}{1-   \frac{c}{1-\alpha}(x^{1-\alpha}-1)}.
 \end{align} It decreases  with the growth of $x$, when $x\leq R_2 ((\alpha+1)/2)^{1/(1-\alpha)}$.


To  find the evolutionary mechanism underlying the early  decreasing dropouts,
we also came back to the mechanism underlying such a power law.
  The durations of learners approximately following a power law   are less than those approximately following  an  exponential distribution. Hence   it is  more reasonable
to regard  their learning behavior   as memorylessness. Therefore, we   supposed the durations of those learners are also  governed  by Eq.~(\ref{eq2}).
 Meanwhile,    power law   reflects the heterogeneity  of samples~\cite{Xie6,Xie8}.
 It means the parameter $\lambda$ of Eq.~(\ref{eq2})  should be  heterogenous over learners.

 We expressed this heterogeneity by   $\lambda(\nu)= {\nu^{b}}/{a}$, and then
 \begin{align}\label{eq8}
  \frac{d }{d t}k_i(t)= \frac{ {\nu^{b}} }{at},
 \end{align}
   where $\nu$ is a random integer  of the uniform distribution over $[S_1, S_2]$, $a>0$, and $b>1$.
Solving it on interval $[y,T]$ gives rise to
   \begin{align}\label{eq9}  k_i(T)  =  {\frac{\nu^{b}}{a}}\log{\frac{T} {y} },  \end{align}
   where     $y$ is  the start time of viewing, sampled from the uniform distribution over $[T_0, T]$.
   Hence  the
 expected value of
    $k_i(T)$ is   $  { \nu^{b} \log 2}/a    $, which   yields
  \begin{align}\label{eq10}  p(k_i(T)\leq x)
 =&p\left(\nu \leq {\left( \frac{ax}{\log2} \right)}^{  \frac{1} {b}}   \right ) \notag\\  =&
  \frac{1}{(S_2-S_1+1)}
  { \left(\frac{ax}{\log2}\right)}^{  \frac{1} {b}} . \end{align} Then
the density function of viewing duration  is
 \begin{align}\label{eq11} p(k_i(T)=x)=&\frac {d}{dx}{p(k_i(T)\leq x)}
  \propto  x^{  \frac{1}{b}-1}. \end{align}

The strict   deduction of the density function
 needs     averaging over all possible $\nu$, which yields
\begin{align}\label{eq12}
p(k_i(T)=x)=&\frac{1}{S_2-S_1}\int^{S_2}_{S_1} \frac{1}{\lambda(\nu)}\mathrm{e}^{-\frac{1}{\lambda(\nu)}x}d\nu\notag\\
  = & \frac{1}{S_2-S_1}\int^{S_2}_{S_1}a \nu^{-b} \mathrm{e}^{-a \nu^{-b} x}d\nu  \notag\\
=&\frac{a^{ \frac{1} {b}}}{ b(S_2-S_1)}   {x}^{ \frac{1}{b}-1}   \int^{aS^{-b}_1  x}_{ aS^{-b}_2x}\tau^{-\frac{1}{b}} \mathrm{e}^{-\tau}d\tau \notag\\
\propto &      {x}^{ \frac{1}{b}-1} I(x),\end{align}
where $I(x)=  \int^{a S^{-b}_1 x}_{a S^{-b}_2x}\tau^{-\frac{1}{b}} \mathrm{e}^{-\tau}d\tau$.
Differentiate the integration  part to obtain
\begin{align}\label{eq13}
\frac{d}{dx} I(x)=
a^{1-\frac{1}{b}} x^{-\frac{1}{b}}\left(
S^{1-b}_1\mathrm{e}^{-aS^{-b}_1  x}-S^{1-b}_2\mathrm{e}^{-aS^{-b}_2x}
\right).
\end{align}
This derivative is
approximately equal to $0$ if $a$ is large enough, which is guaranteed by the empirical values of $a$ in Table~\ref{tab3}. Hence the integration part is free of $x$  and $p(k_i(T)=x)  \propto  {x}^{  {1}/{b}-1}$.

To make the simulated distributions fit the empirical ones,  we valued
the parameters   of Eq.~(\ref{eq9}) based on empirical data as follows.
  Calculate  the domain   of the durations of segment-learners (which
are   less than one minute) $[R_1,R_2]$, and
fit  their distribution  by power law   $cx^{-\alpha}$.
The fitted values of $\alpha$ and  $c$ are listed   in Table~\ref{tab2} and Table~\ref{tab3} respectively.
  Comparing   the coefficients of Eq.~(\ref{eq11}) to the  $\alpha$  and $c$    gives rise to
 $\alpha=1- {1}/{b}$ and $c=   { \left( {a} /{\log2}\right)}^{   {1}/ {b}}/{(S_2-S_1+1)b} $.
 Solving them  obtains
the value of $a$ and $b$.
The
 expected duration
      $  { \nu^{b} \log 2}/a$ belonging to  $[R_1, R_2]
  $  gives rise to the sampling  interval   for  $\nu$ and then for  $y/T_2$. Table~\ref{tab7} in the appendix shows
the detail for this simulation process.


Above analysis realizes a process of deriving   power law   from  a range of exponential distributions. Moreover, it provides an
explanation for the early decreasing trend of the hazard functions. That is, the
dropout rate $1/\lambda(\nu)$ decreases with the growth of the
expected value $\lambda(\nu)$.
 The simulations based on the
solution~(\ref{eq9}) also provide a reasonable fit to the heads    of the empirical duration distributions (Fig.~\ref{fig5}).
 Therefore, Eq.~(\ref{eq8}) can be regarded as an expression of the evolutionary mechanism for  power law and for the early decreasing trend. In addition,
the memorylessness of   Eq.~(\ref{eq8}) together with that of Eq.~(\ref{eq3}) can be regarded as the  intrinsic meaning of
   the class name    segment-learners.

\section*{Discussion and conclusions}
\begin{figure*}\centering
\includegraphics[height=3.   in,width=4   in,angle=0]{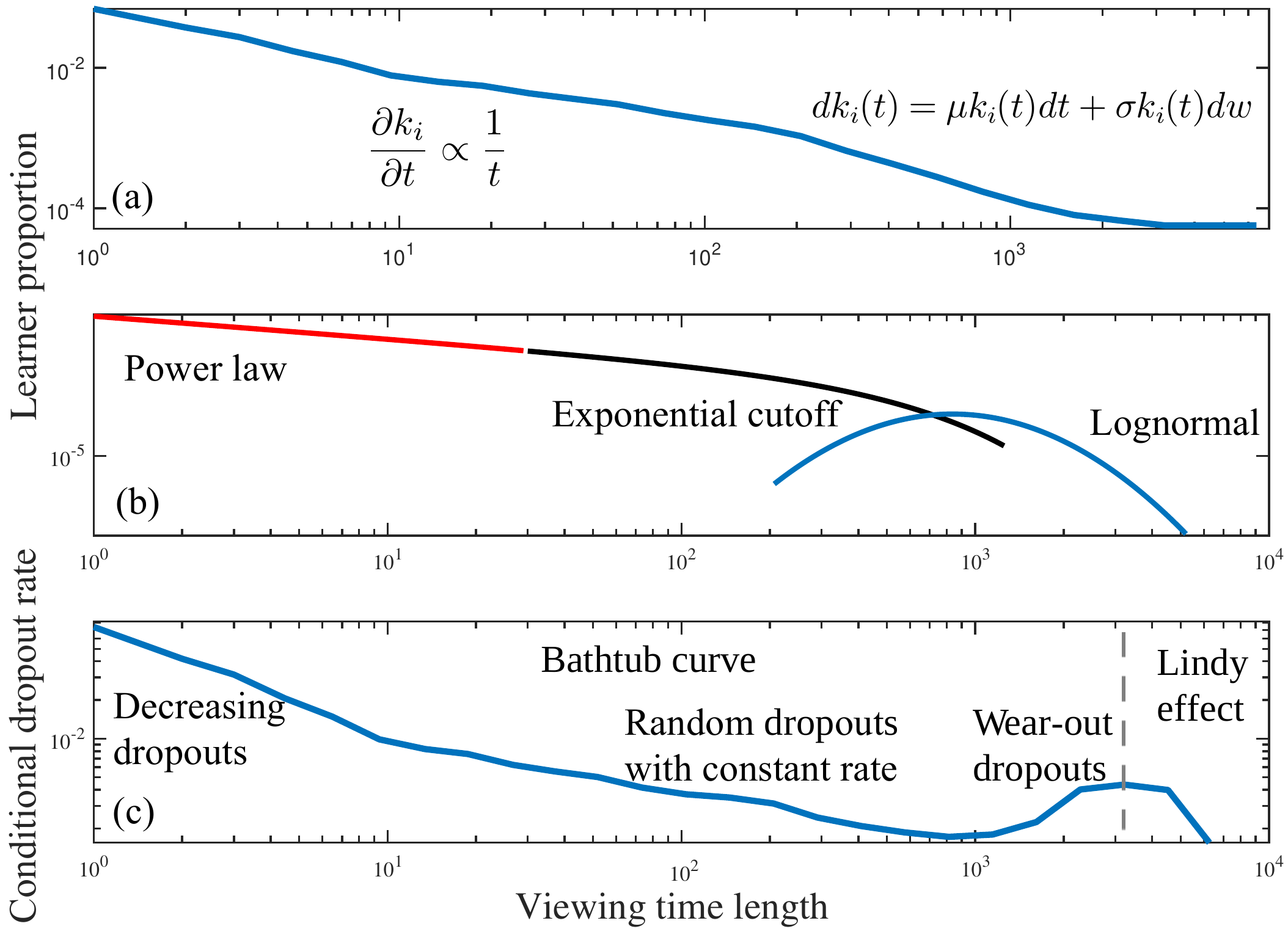}
\caption{  { \bf  An illustration  of the presented results. }
The illustrated data  are from the course {\it C~language}. Panel (a) shows the density function   and the evolutionary equations for viewing durations. Panels (b, c) show the features of the density function, and those of the corresponding hazard function. The equation on the left describes the evolutionary mechanism for power law with exponential cutoff,   early deceasing dropouts, and  the random dropouts with a constant rate. The equation on the right describes the mechanism for lognormal distribution,   wear-out dropouts, and   the Lindy effect.
  } \label{fig7}      
\end{figure*}

\noindent
The survival analysis on the  viewing behavior  of learning MOOCs shows
 the hazard functions of empirical viewing   durations are featured by     the  Lindy effect and bathtub curve  simultaneously.
   Two random differential equations are provided to describe the growth processes of viewing durations.
  The solutions to these equations
   predict the features of the hazard functions.
 Therefore,   these equations   can be regarded as   mathematical  expressions  of the  evolutionary  mechanisms underlying these features.
We summarized the  presented results in Fig.\ref{fig7}.



\begin{figure*}\centering
\includegraphics[height=1.7   in,width=6   in,angle=0]{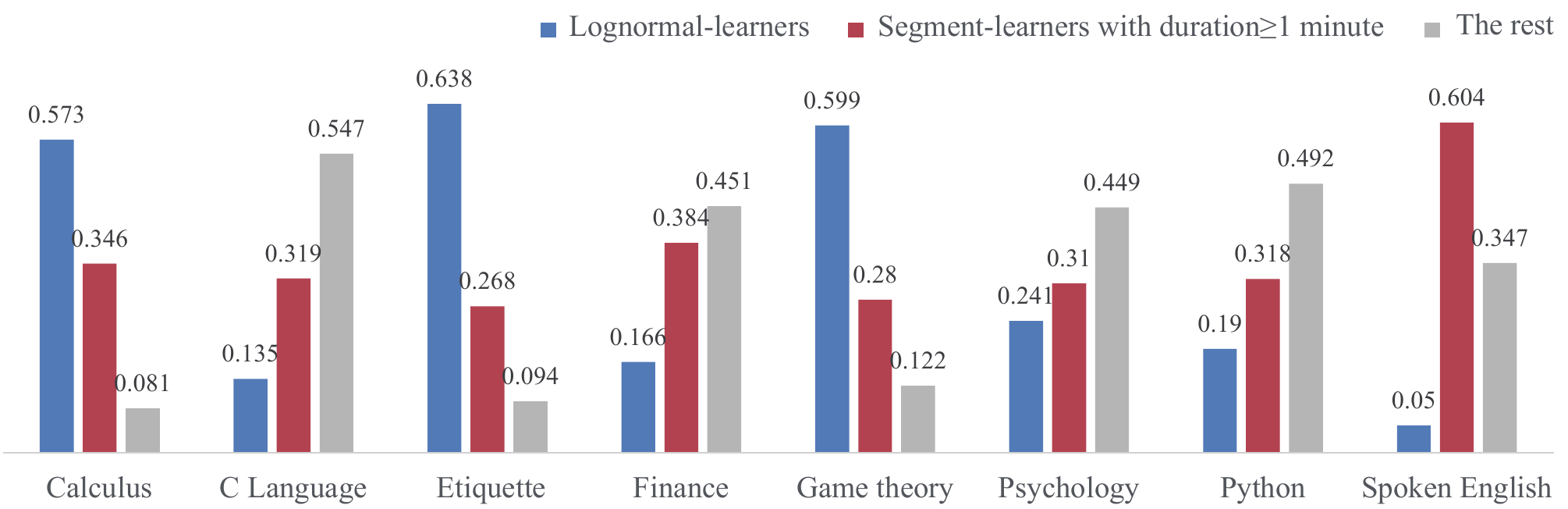}
\caption{  { \bf  The fractions   of three learner-types. }
The three types of learners are
lognormal-learners (the durations of them follow  a lognormal distribution), the segment-learners with duration no less than one minute (those durations follow
an exponential distribution approximately), and the rest (the durations of them    follow
a power law approximately).  } \label{fig8}      
\end{figure*}

 The presented results
have the potential to illuminate specific views and implications in a broader
study of learning  behaviors.
For example, the features of  viewing duration distributions can be used to profile  the type of learners, such as  lognormal-learners, the learners with a duration approximately
following an exponential distribution, and those with a  duration  approximately following a power law.
The fractions   of these types  vary over courses~(Fig.~\ref{fig8}). Over half     of the learners studying the  course {\it Calculus}    are   lognormal-learners.
Almost half of the learners taking
{\it C Language} or {\it Python}  viewed less than one minute, where their  duration   approximately  follows  a power law.
Weighting   each type with a different value  helps to   measure  the attractions of MOOCs in a reasonable way.

Comparing  the  duration distributions   before and after adopting a teaching   method helps to know whether
the method   significantly
increases   or decreases    learning durations. For example,
if the KS test shows  the  duration distributions of lognormal-learners are identical, it cannot say the improvement of the adopted method is significant.
 This   can also be used to compare the attractions of different courses.
 It removes the heterogeneity of  the number of learners, and hence is  a fair way for  the  courses  with high quality  but having   few learners.

In the presented  equations, the   viewing durations are  based on  random factors, and  memory or  memorylessness.
   At the beginning of studying a course,  a learner does not need the knowledge of the course. When  studying  deeply, the learner would need  the
 knowledge   learned from the  course.
 This process  could  be regarded  as the  transition from memorylessness  to memory.
 Meanwhile,  the viewing duration distribution  of each empirical course has a fat tail,
   known as a feature  of   complexity.
We found that each  tail
   is dominated by   the tail  of a lognormal distribution, and that viewing with memory can generate  a lognormal distribution.
Therefore, exploring the mechanisms underlying  the transition  would contribute to understanding the role of memory in the  complexity of   learning behaviors.

 \section*{Appendixes}
\subsection*{Categorization of learners}
The following    algorithm  comes from  Reference\cite{Xie7}.
\begin{table*}[!ht] \centering \caption{{\bf An  algorithm   of categorizing  learners.} }
\vskip 3mm
\begin{tabular}{l r r r r r r r r r} \hline
Input: the viewing  duration  $t_s$ and   the number of viewed
videos  $n_s$  of learners $L_s$ $(s=1,...,m)$.\\
\hline
For   $k$ from $0$ to $\max(n_1,...,n_m)$: \\
~~~~do the KS test for $t_s$ of the learners $L_s$ satisfying  $n_s> k$ with   the null
 hypothesis that they follow a \\ ~~~~lognormal distribution;\\
~~~~break  if  the test cannot reject  the null hypothesis  at   significance level  $5\%$. \\ \hline
Output:   the current $k$ (denoted as $\kappa$). \\ \hline
 \end{tabular}
   \begin{flushleft}
 The unit of durations is
millisecond.
    Categorize   learner    $L_s$    as a  lognormal-learner if    $n_s>\kappa$,   and  as  a segment-learner if else.
   \end{flushleft}
\label{tab4}
\end{table*}

\subsection*{The process  of generating synthetic durations}

\begin{table*}[h] \centering \caption{{\bf Modelling   exponential cutoff.} }
\vskip 3mm
\begin{tabular}{l r r r r r r r r r} \hline
Input:  the   durations ($\geq 1$ minute) of  segment-learners.\\
\hline
Regress the coefficients of $\delta\mathrm{e}^{-x/\lambda}$  based on the input.\\
Calculate the input's domain   $[T_1, T_2]$.\\
For $i$ in range $1$ to the number of  empirical durations:\\
~~~~sample a $y/T$ from  the uniform distribution over $[\mathrm{e}^{-T_2/\lambda}, \mathrm{e}^{-T_1/\lambda} ]$;\\
~~~~substitute it into Eq.~(\ref{eq4}) to obtain a random integer;\\
~~~~append the   integer   to the list of synthetic durations.  \\
\hline
Output:  the list of synthetic durations. \\ \hline
 \end{tabular}
    \begin{flushleft}
The unit of   durations is  2 seconds.
    \end{flushleft}
\label{tab6}
\end{table*}

\begin{table*}[h] \centering \caption{{\bf Modelling   power-law part.} }
\vskip 3mm
\begin{tabular}{l r r r r r r r r r} \hline
Input:  the   durations ($< 1$ minute) of  segment-learners; the domain $[S_1,S_2]$ of $\nu$.\\
\hline
Regress the coefficients of $cx^{\alpha}$ based on the input.\\
Calculate the parameters of Eq.~(\ref{eq9}):
  $b=1/(1-\alpha)$,  $a= \left(c(S_2-S_1+1)b\right)^b \log2   $.
 \\
Calculate  the input's domain $[R_1, R_2]$.\\
For $i$ in range $1$ to  the number of  empirical durations:\\
~~~~sample a $\nu$ from  the uniform distribution over  $[(aR_1/\log2)^{1/b}, (aR_2/\log2)^{1/b}   ]$;\\
~~~~sample a $y/T_2$ from   the uniform distribution over  $[\mathrm{e}^{-R_2/\lambda(\nu)}, \mathrm{e}^{-R_1/\lambda(\nu)} ]$;\\
~~~~substitute them into Eq.~(\ref{eq9}) to obtain a random integer;\\
~~~~append the integer to the list of synthetic durations.  \\
\hline
Output:  the list of synthetic durations. \\ \hline
 \end{tabular}
   \begin{flushleft}
The unit of   durations is  2 seconds,     $S_1=1$, and $S_2=29$.
   \end{flushleft}
\label{tab7}
\end{table*}



\section*{Acknowledgments}
The author thinks Researcher
Xiao Xiao in the Higher Education Press, Professor Ming Zhang in the Peking university, Professor  Jinying Su and Jianping Li in the National University of Defense Technology for their helpful comments and feedback.
The author is grateful to
the MOOC platform iCourse for its empirical data. This work is supported by    National   Science Foundation of China (Grant No. 61773020).

\end{document}